\documentclass[journal]{IEEEtran} 
\IEEEoverridecommandlockouts
\usepackage{cite}
\usepackage{lipsum}
\usepackage{flushend}
\usepackage[dvips]{color}
\usepackage[colorlinks=false]{hyperref}
\usepackage{epsf}
\usepackage{epsfig}
\usepackage{graphicx}
\usepackage{graphics}
\usepackage{epstopdf}
\usepackage[footnotesize]{caption}
\usepackage[bottom]{footmisc}
\usepackage{amsmath}
\usepackage{amssymb}
\usepackage{amsthm}
\usepackage{amsxtra}
\usepackage{algorithm}
\usepackage{algorithmic} 	
\usepackage{enumerate}
\usepackage{multirow}
\usepackage{enumerate}
\usepackage{amssymb}
\usepackage{lipsum}
\usepackage{setspace}
\usepackage{multirow}
\usepackage{tikz}
\usepackage{setspace}
\usepackage{subfigure}

\newtheorem{remark}{Remark}

\usepackage{rotating} 
\usepackage{graphicx} 
\makeatother
\begin{document}
\title{Smart Grid Testbed for Demand Focused Energy Management in End User Environments}
\author{\IEEEauthorblockN{Wayes Tushar, Chau Yuen, Bo Chai, Shisheng Huang, Kristin L. Wood, See Gim Kerk and Zaiyue Yang}
\thanks{W. Tushar, C. Yuen and K. L. Wood are with the Singapore University of Technology and Design (SUTD), Singapore.}
\thanks{B. Chai is with State Grid Global Energy Interconnection Research Institute, Beijing, 102211, China.}
\thanks{S. Huang is with the Ministry of Home Affairs, Singapore.}
\thanks{S. G. Kerk is with the Power Automation, Singapore.}
\thanks{Z. Yang is with Zhejiang University, China.}
}
\IEEEoverridecommandlockouts
\maketitle
\begin{abstract}
Successful deployment of smart grids necessitates experimental validities of their state-of-the-art designs in two-way communications, real-time demand response and monitoring of consumers' energy usage behavior. The objective is to observe consumers' energy usage pattern and exploit this information to assist the grid in designing incentives, energy management mechanisms, and real-time demand response protocols; so as help the grid achieving lower costs and improve energy supply stability. Further, by feeding the observed information back to the consumers instantaneously, it is also possible to promote energy efficient behavior among the users. To this end, this paper performs a literature survey on smart grid testbeds around the world, and presents the main accomplishments towards realizing a smart grid testbed at the Singapore University of Technology and Design (SUTD). The testbed is able to monitor, analyze and evaluate smart grid communication network design and control mechanisms, and test the suitability of various communications networks for both residential and commercial buildings. The testbeds are deployed within the SUTD student dormitories and the main university campus to monitor and record end-user energy consumption in real-time, which will enable us to design incentives, control algorithms and real-time demand response schemes. The testbed also provides an effective channel to evaluate the needs on communication networks to support various smart grid applications. In addition, our initial results demonstrate that our testbed can provide an effective platform to identify energy wastage, and prompt the needs of a secure communications channel as the energy usage pattern can provide privacy related information on individual user. 
\end{abstract}
\section{Introduction}\label{sec:introduction}
The smart grid is a power network composed of intelligent nodes that can operate, communicate, and interact autonomously to efficiently deliver electricity to their consumers. It features ubiquitous interconnections of power equipments to enable two-way flow of information and electricity so as to shape the demand in order to balance the supply and demand in real-time. Such pervasive equipment interconnections necessitate a full-fledge communication infrastructure to leverage a fast, accurate, and reliable information flow in the smart grid. In this context, the research on different aspects of smart grid has gained significant attention in the past few years, e.g., the literatures surveyed in~\cite{Fang-J-CST:2012}. Although a lot has been done from theoretical perspectives, it is until recently when the actual implementation of prototypes has been given deliberate consideration. Currently, a considerable number of research groups are working towards establishing testbeds to validate designs and implemented protocols related to smart grid. These testbeds have various aims, scale, limitations and features; these have been summarized and presented in Table~\ref{table:1}.

Most of these testbeds conduct experiments either in a lab environment (e.g., SmartGridLab, VAST, Micro Grid Lab, cyber-physical), or in isolation, in a residential (e.g.,PowerMatching City) or a commercial space (e.g., Smart Microgrid). In the testbeds surveyed, only the JEJU testbed is comprehensive enough to consider both the residential and commercial paradigm with user and grid interactions. Moreover, in spite of considerable on-going studies towards smart grid prototypes, and the massive efforts by utilities and local authorities, deployment of smart grids has met with near customer rejection\footnote{http://www.greentechmedia.com/articles/read/illinois-rejects-amerens-smart-grid-plans1.}. Therefore, there is much impetus on designing and implementing well accepted solutions to bolster consumer-grid interactions.
 
In this respect, this paper presents the main accomplishments towards a user-centric smart grid testbed design at the Singapore University of Technology and Design (SUTD). The highlight of the paper is mainly on the communication infrastructure that has been implemented in the testbed in order to provide ICT services to support a greener smart grid. The testbed is simulated in an approximate real world scenario, where the student dormitory (a 3 bedroom unit with 6 to 9 resident students) is the approximated residential space, and the faculty offices and shared meeting rooms are proxy commercial office spaces. The study focuses on both residential and commercial consumers and their interaction with a central administrative body, e.g., the grid or the intelligent energy system (IES), where the IES is an entity that provides alternative energy management services to the consumers. 

The testbed at SUTD consists of two networks: a home area network (HAN) and a neighborhood area network (NAN). A HAN network is implemented within each residential unit or at SUTD office in order to collect different energy related data via a unified home gateway (UHG). A NAN, on the contrary, is developed to connect each HAN to the data concentrator through various communication protocols. We provide further detail discussion on both HAN and NAN networks that are implemented at SUTD in Section~\ref{sec:section-3}.

Due to the interactive nature of the system, an efficient and reliable communication infrastructure of low delay is very important. Therefore, this paper mainly discusses the various communication aspects such as broadband power line communication cables (BPL), TV white space (TVWS), HAN and NAN of the testbed at SUTD. Benefits that could be attributed and examined in this system include lowering of operational costs, increasing ancillary electricity options and better peak demand management. We would also design and examine suitable incentives, which would encourage consumers to adopt these demand response mechanisms for participation in the market by exploiting the use of the testbed.
\section{Testbed Components, Features and Challenges}\label{sec:challenges}
\begin{table*}
\caption{Attributes of Different Smart Grid Testbeds.}
\includegraphics[width=\textwidth]{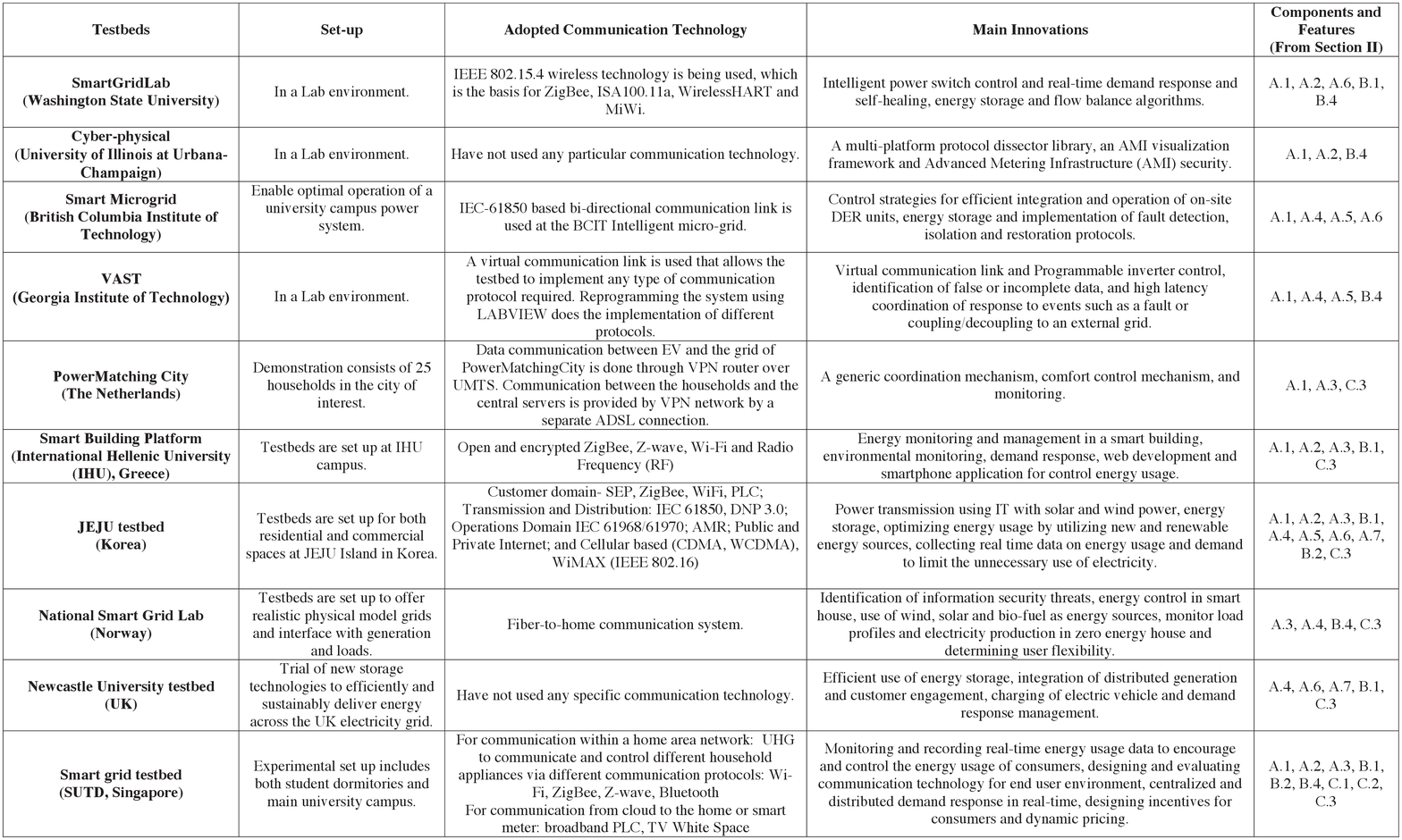}
\label{table:1}
\end{table*}
The smart grid is envisioned as a future power network with hundreds of millions of  endpoints, where each endpoint may generate, sense, compute, communicate and actuate. Deployment of such complex systems necessitates sophisticated validation prior installation, e.g., by establishing smart grid testbeds. However, the rapid fluctuation in power supply and demand, voltage and frequency, and active consumers' interaction with the grid make the practical realization of testbeds extremely challenging. To this end, we classify the suitable technologies and features of such testbeds into three categories: 1) Hardware based components, 2) Software based elements, and 3) Other features. We give a brief description of different technologies and elements within each of the defined categories and their impact on the design of a communication network as follows.
\subsection{Hardware based components}
\subsubsection{Two-way communication network}In a smart grid, a large number of sensors, actuators, and communication devices are deployed at generators, transformers, storage devices, electric vehicles, smart appliances, along power lines and in the distributed energy resources. The optimal control and management of these nodes in real-time is essential for the successful deployment of smart grids~\cite{Maharjan-TSG:2013}, which gives rise to the need for fast and reliable two-way communication infrastructures. For example, depending on the online requirements to response, reserve power is required in the smart grid in case of an unexpected outage of a scheduled energy resource, and required response has to be deployed within $30$ seconds\footnote{Reserve power can be split into primary (less than $30$ seconds), secondary (less than $15$ minutes) and tertiary ($15$ minutes), depending on required response times to monitor and control.} . 
\begin{figure*}[t]
\centering
\includegraphics[width=\textwidth]{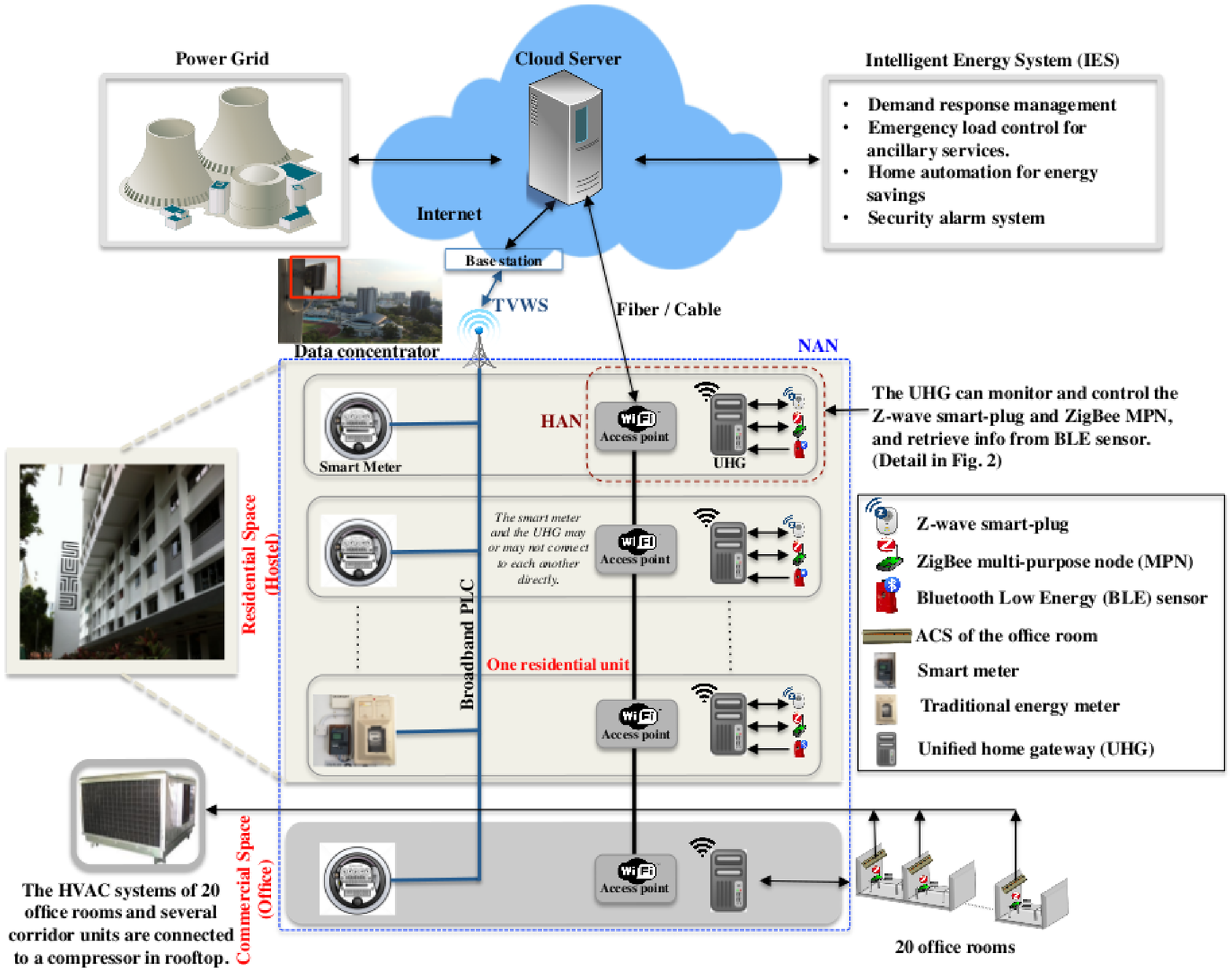}
\caption{Overview of SUTD testbed consisting of NAN and HAN.} \label{fig:NAN}
\end{figure*}
\subsubsection{Advanced metering infrastructure (AMI)}The AMI measures, collects and analyzes the energy usage, and communicates with metering devices either on requests or on a schedule. There is an increasing concern regarding their costs, security and privacy effects, and the remote controllable ``kill switch" included in them. A scalable secured MAC protocol becomes critical for the AMI system.
\subsubsection{HAN and NAN} In accordance with \cite{Rong-Networks:2011}, we organize the smart grid network into a HAN and NAN. A HAN is the core component of establishing a residential energy management system as part of a demand response management (DRM) scheme in a smart grid. Establishing the HAN is required to maintain an internet-of-things protocol for different sensors/actuators that run on different physical layer communication protocols within a home. This consequently imposes the necessity of devising gateways/controllers/central communicators, which are capable of  handling these miscellaneous communication techniques. Depending on the business model, there could be multiple NANs. For example, AMI is setup and maintained by the grid, and is connected to one NAN; while third party IES (e.g. Google NEST) could be on another NAN that provides additional services such as home automation, energy management, or security services. 
\subsubsection{Renewables supply} The generation from renewable sources such as wind and solar is intermittent and unpredictable~\cite{Wayes_ITS:2016,Naveed_Access:2015}. Hence, how to design the control architectures, algorithms, and market mechanisms to balance supply and demand, support voltage, and regulate frequency in real-time are the key challenges in the presence of these volatile green energy supplies.
\subsubsection{Smart power electronics}Smart components such as DC/AC inverters, flexible AC transmission systems and smart switches will be increasingly deployed on various components such as the transmission systems, transformers and power generators in the smart grid. Hence, there are needs of solutions that will enable the design of power electronics and control appliances to maximize reliability, efficiency and cost reduction for the power grid.
\subsubsection{Energy storage system}An energy storage system stores electricity when the demand is low and provides the grid/IES with electricity when the demand is high. Energy storage systems are becoming an essential part of smart grids  due to the increasing integration of intermittent renewables and the development of micro grids~\cite{Atzeni-TSG:2013}. Innovative and user-centric solutions are required to determine the type and size of energy storage according to the application, to determine the optimal storage location and reduce the cost of storage devices.
\subsubsection{Electric vehicles (EVs)}EVs can act as mobile storage devices and assist the grid to fulfil various energy management objectives through grid-to-vehicle (G2V) and vehicle-to-grid (V2G) settings. Key challenges are: addressing issues like quick response time, power sharing, EV charging protocol design and exploiting the use of EVs storage devices for power compensation~\cite{Wayes_ITS:2016}, and voltage and frequency regulation services~\cite{huang2012effects}.
\subsection{Software based elements}
\subsubsection{DRM}The two-way flow of information and electricity in smart grids establishes the foundation of DRM; which is in fact the change of electricity usage patterns by the end-users in response to the incentives or changes in price of electricity over time. DRM can be accomplished in either a centralized or a distributed fashion \cite{Hassan-Energies:2013}. A distributed DRM algorithm greatly reduces the amount of information that has to be transmitted when compared to a centralized DRM algorithm. Therefore, a smart grid testbed needs to have the capability to investigate both DRM schemes, if possible, in order to examine their effectiveness on the target load management.
\subsubsection{Dynamic pricing} Dynamic pricing is critical for effective DRM in smart grids as a driver for behavior change~\cite{TaoJiang_TSG:2014}. The pricing scheme needs to be beneficial to the grid in terms of reduction in operational cost, energy peak shaving and valley filling. It also needs be economically attractive to consumers by reducing their electricity bills and should not cause significant inconvenience to them for changing their energy consumption behavior, i.e., for scheduling or throttling. Securing the pricing information is important to protect against malicious attacks, e.g., an outdated pricing information is injected to destabilize the smart grid.
\subsubsection{Outage management}In a time of crisis, the capacity for fault detection and self-healing with minimal restoration time and improved efficiency, i.e., outage management~\cite{LiangYu_TPDS:2015} is important. Hence, with real-time insight, the smart grid testbed should possess automatic key processes to easily locate and route power around affected spots to reduce unnecessary truck rolls and save costs.
\subsubsection{Security and privacy}Cyber security is a pivotal challenge for smart grid. In fact, cyber-based threats to critical infrastructure are real and increasing in frequency. However, the testing of potential threats are challenging due to the general lack of defined methodologies and prescribed ways to quantify security combined with the constantly evolving threat landscape. Moreover, as the participation of consumers in smart grid becomes more prevalent, privacy information, e.g., their home occupancy, will be more vulnerable. Hence, applied methods and rigorous privacy and security assessments are necessary for complex systems where heterogeneous components and their users regularly interact with each other and the IES.
\subsection{Other features}
\subsubsection{Ancillary Services}Regulation power allows for the grid to manage second-to-second fluctuations from forecasted demand and reserve markets enable electricity providers to have instant-on solution to kick in when power delivery problems emerge. Theoretically, smart grids are able to provide such services through DRM~\cite{Tushar-TIE:2015}. Therefore, smart grid solutions need to enable the grid providers to combine the DRM schemes  with energy storage solutions to make the grid system more reliable.
\subsubsection{Scalability} A smart grid system may consists of millions of active nodes, which need to be managed in real-time. Accordingly, optimal economic mechanisms and business models are required for engendering desired global outcomes. However, it is extremely difficult to maintain such a large scale and complex cyber-physical system in real-time. Besides, features like automatic fault detection, self-healing, autonomous distributed demand-side management and disaster management make the conservation more strenuous. Hence, innovations are required to improve the scalability of such a huge system without affecting any of its features. For instance, instead of using different gateways for communicating with different equipment in a home (where each devices may run on different communication protocols), devising a single UHG could make a HAN considerably scalable. A scalable MAC protocol would be interesting to explore in dense residential areas, where over thousands of machine-type devices such as smart meters communicate at the same time~\cite{Liu-JIEEEIT:2014}.
\subsubsection{User acceptability and participation} One of the key challenges for the success of smart grids is to be able to clearly elucidate the benefits for the consumers~\cite{Tushar-TSG:2014}. One solution could be a clear user friendly interface (e.g., smart phone apps) to demonstrate the ability to visualise electricity usage and real-time electricity prices, so as to ease energy savings for the consumers through smarter load shifting. 

It can be noticed that advanced ICT technologies and communications networks play important roles in various applications, features, and even in the acceptance of a smart grid. To this end, we now discuss the main features of the testbed at SUTD in the next section.
\section{SUTD Testbed Overview}\label{sec:section-3}
The aims of the SUTD smart grid testbed are to develop innovative technical approaches and incentives for smart energy management, through the integration of commercial space strategies, residential space strategies, and the design and development of supporting technologies. Hence, testing the  suitability of various communications technologies for smart grid so as to address some of the challenges mentioned in Section \ref{sec:challenges} is especially critical. The interaction between the consumers and the grid/IES has been prioritized in this work, which leads us to divide the whole smart grid testbed into two networks: 1) NAN and 2) HAN. The IES set up the HAN, which is connected to the cloud directly though broadband internet, while the grid will set up the NAN that connects the smart meters. However, if the IES is provided by the grid, then the HAN and the UHG would certainly be connected under the same network. To keep the structure general, we consider them to be separated in the SUTD testbed. To this end, what follows are the brief descriptions of the NAN and HAN, and the associated technologies used in the SUTD testbed.

\subsection{NAN}
The NAN of the testbed is composed of a number of HAN units from either the SUTD dormitories or the university main campus and the cloud. In our setup, we employ two technologies for NAN: one is BPL and TVWS for smart meters setup by the grid, and another is the traditional fiber / cable internet, where all the UHGs are connected via Wi-Fi access points, which could be setup by a 3rd party IES. In Fig.~\ref{fig:NAN}, we show a schematic diagram of how the NAN is set up within the SUTD testbed. Now, we provide a brief description of the used communications technologies, i.e., BPL and TVWS, in the following subsections.
\subsubsection{BPL}In the testbed, the communication between the smart meter outside each residential unit and the data concentrator at the top of the apartment building  is based on BPL. This choice of communication between them is simple and obvious as no additional cabling is needed in connecting all the units within an apartment block. It provides superior performance across thick walls as compared to any other types of communications and simplified the entire network management of smart meters.
\subsubsection{TVWS}At SUTD, TVWS is used for linking the data concentrator from every apartment block to the base-station before data is uploaded to the cloud. In Table \ref{table:tvws} we show the detail specification of TVWS that we have used at SUTD. Currently, Infocomm Development Authority (IDA) is setting up trials and the standardization of TVWS in Singapore. Please note that TVWS is a freely available spectrum and thus provides a great opportunity for many new small and virtual operators to setup their networks for M2M applications at low cost. In this testbed, building a NAN on TVWS spectrum also provides more operational flexibility that can easily be scaled up by adding additional Base Stations (BSs). Therefore, for NAN wireless access provides good solution and wide coverage without much installation cost. Furthermore, TVWS (unlicensed band) can greatly reduce the cost, as compared to licensed band and fixed access.

Our testbed based on BPL and TVWS provides a dedicated and separated network from the Internet cabling, which is also used for web surfing or video streaming. It provides an opportunity to test the network delay and reliability for smart grid applications, and can be compared against the HAN that will be discussed next. 

\begin{table}[t]
\centering
\caption{The technical specifications of TVWS network deployed at SUTD campus.} \label{table:tvws}
\includegraphics[width=\columnwidth]{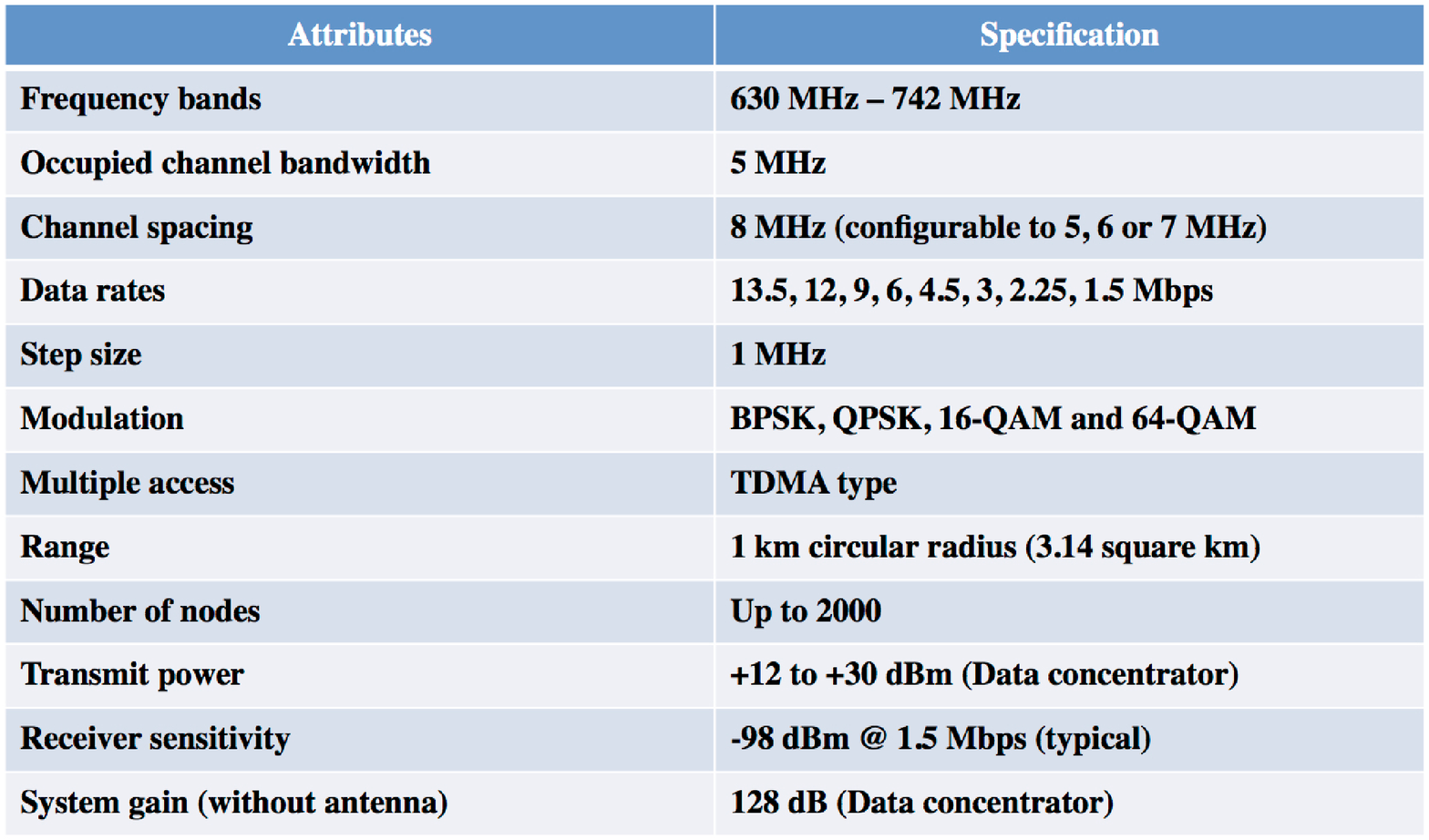}
\end{table}

\begin{figure*}[t]
\centering
\includegraphics[width=\textwidth]{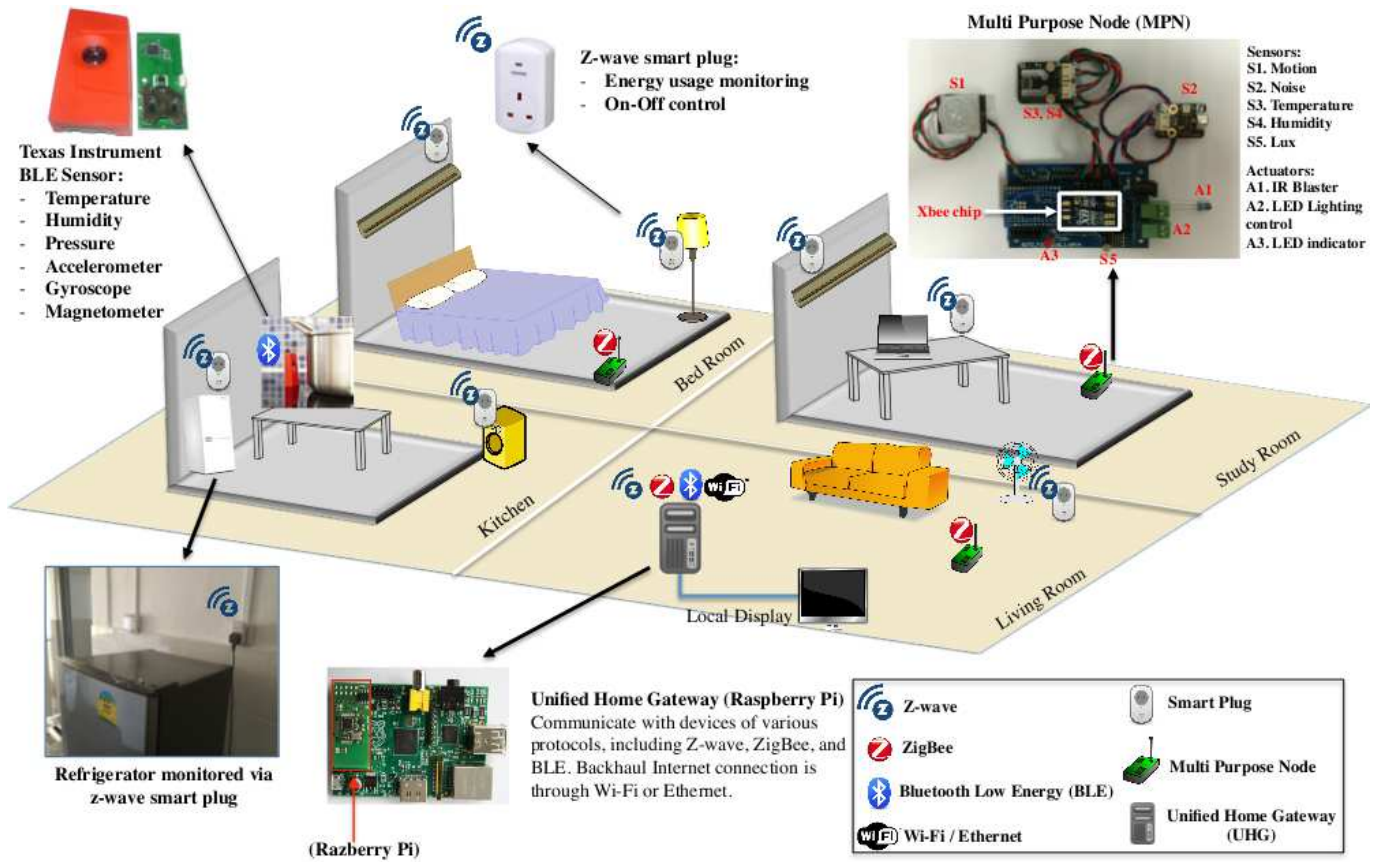}
\caption{Home Area Network.} \label{fig:HAN}
\end{figure*}
\subsection{HAN}
While all participating units from the dormitories and the campus form the NAN together, each unit is equipped with its own HAN inside the house as shown in Fig.~\ref{fig:HAN}. A UHG is the central point of attention of a HAN, through which the IES can monitor the energy usage pattern or control each of the equipments of the unit within the network. Here, on one hand, the gateway adopts two-way communication protocols such as ZigBee, Z-wave and Bluetooth to connect with equipments and sensors in the house. On the other hand, it is connected to the IES through Wi-Fi for sending the monitored information to the IES.
\subsubsection{UHG} A novel aspect of the SUTD testbed is the introduction of a UHG in each of the HAN unit of both residential and commercial space areas, built on a Raspberry Pi computer (Model-B Rev 1) as shown in Fig.~\ref{fig:HAN}. Please note that the HAN network consists of different type of smart plug and sensors that may need to communicate with each other. In this context, Universal home gateway (UHG) is developed and used in the testbed to facilitate such heterogeneous communication between different sensors and smart plugs. The UHG is capable of handling a number of communication protocols including Z-wave, ZigBee and Bluetooth for communicating with different smart devices such as smart plugs and various sensors. The UHG directly transmits the monitored data from a HAN unit to the IES through Wi-Fi. The challenge of such a UHG is that it needs to communicate with devices (e.g. smart-plug / sensors) of various protocols. The delay between the commands from the IES to the UHG and from the UHG to the end devices could also pose challenges depending on the applications. It however provides a scalable solution that enables thousands or even ten thousands of households in the system. In the SUTD testbed, we have implemented Restful HTTP and XMPP protocol for the communication between the UHG and the cloud of the IES. Hence, the UHG serves as a gateway for many non-Internet Protocol based sensors to connect to the Internet.

\subsubsection{ZigBee} ZigBee is a low-cost and low-power wireless mesh network specification, which is built on the top of the IEEE 802.15.4 standard for low-rate wireless personal area networks (WPANs), and operates in the ISM frequency band 2.4 GHz. ZigBee is essentially a non-IP based communication protocol and thus suitable for communication in testbeds set up in a non-residential setting. This is due to the fact that in SUTD campus (which is also the same as many other commercial/industrial network), the Wi-Fi network has very strict security, where the password is changed on regular basis (e.g., in every three months at SUTD campus). However, it is significantly difficult to change the password of every sensor in every three months time. Therefore, non-IP based solution is required and we use ZigBee in the MPN to serve that purpose. Besides, for a multi-hop extension considering the office scenario, we require a protocol that can support multi-hop to extend the coverage to reduce the needs of a gateway (within a house, however, single hope with UHG is usually good enough). Furthermore, ZigBee provides security (e.g. the ZigBee provides basic security where only nodes of the same network id can join in the network) and the simplicity of having two-way communication facilities.

The multi-purpose node (MPN) designed in SUTD, as shown in Fig.~\ref{fig:HAN}, consists of a number of low power sensors and actuators. It is equipped with a motion detector, a temperature and humidity sensor, a noise sensor and a Lux sensor. It communicates with the UHG through its Xbee Chip, which is a ZigBee communication module, to send the monitored information. The MPN is capable of controlling devices through its actuators like an IR Blaster (to control the air conditioning system (ACS)) and potentiometer (to control the LED light power supply). The MPN is driven by an Arduino Fio micro controller, which is installed within the MPN. In every hostel unit in SUTD, each HAN consists of 4 MPNs, one for each bedroom, and one for the living room. While in the office, there are 20 MPNs (one for each faculty office) in one section of an office block connected through multiple ZigBee relays to a single UHG. The UHG also fulfils an important role later on, when distributed DRM is implemented, such that users occupancy and electric appliance usage info can be stored locally without sending to the IES to protect users' privacy. 
\subsubsection{Z-wave} Z-wave is a wireless communication protocol around $900$ MHz that uses a low power technology for home automation. For instance, to monitor and remotely control the energy consumption of different equipments in the HAN, we have used a Z-wave smart-plug for each of the devices in the unit. In the current set-up, electric appliances to be monitored are connected to power outlets through Z-wave smart-plugs, whereby each smart-plug is connected to the UHG through a RaZberry module. The plug is able to monitor the amount of energy consumed by the connected device in real-time and instantaneously send that information to the UHG via Z-wave communication. Further, it also has a remote actuation capability  e.g., delaying an electric appliances due to peak in energy demand. Note that the switching signals of smart-plugs are initiated through either the IES (for centralized control) or the UHG (for distributed control).
\subsubsection{Bluetooth} We use a Bluetooth Low Energy (BLE) sensor based on Texas Instrument's (TI's) CC2541 MCU to monitor the temperature, humidity and pressure level of its surroundings, and to communicate with the UHG via bluetooth. The BLE sensor is also equipped with an accelerometer, a gyroscope and a magnetometer that provide the UHG with information like acceleration, orientation and magnetization of an object respectively. In the course of the experiments, we will distribute such TI sensors among the students so that they can design new Internet-of-Things product that can connect to our UHG and provide new green energy services in smart homes.
\subsubsection{Mobile application} We have developed a mobile application\footnote{The figure is not given due to constraint on the number of figure.} to install in students' and faculties' cellphones to engage them in the energy management experiments at SUTD. For example, they can remotely switch on and off the smart plug or the ACS, or monitor the energy usage collected by the testbed. Dynamic pricing information can also be pushed to them over the application.

\subsection{Challenges}
During the experiments in the testbed, we face a number of challenges including communication delay, subject recruitment for case studies, modeling non-homogeneous scenarios and investigating the mismatch between the theoretical results and the outcomes from the experiments.

First, communication delay is one of the challenges that we have encountered while running the experiments. We understand that there is a trade-off between the sampling rate and communication delay, and the delay can be reduced by increasing the rate of sampling. Delays can also be caused by the time required for data savings and retrieval at the server. For example, if a server needs to retrieve data from large number of sensors at the same time, it may incur some delays~\cite{Cao:2016}. However, one potential way to reduce such delay is to efficiently design the database and use more customized script for increasing the processing speed. Nevertheless, such delay is beyond the scope of this paper. Now considering the response times of primary, secondary and tertiary reserves, our implemented centralized monitor and control scheme can comfortably use for secondary (less than 15 minutes) and tertiary reserve (15 minutes). However, it might not be suitable for primary reserve requiring very fast response time, e.g., less than one minute. Implementing distributed control instead of centralized and thus enable each node under observation to respond very fast, if necessary, could be a potential way to resolve this issue. 

Second challenge is to recruit the subjects for experiments who will allow us to install sensors and communication modules in their rooms for monitoring and control purposes. Although the experiments are completely Institutional Review Board (IRB) approved and are designed to conduct in an academic environment, it is hard to find participants for the experiments. This is due to the fact that the participants are concerned about their privacy and worry about the inconvenience that may create due to energy management. Further, students lifestyle is significantly different from the lifestyle of typical residents. In this context, as a token of encouragement for participation, we have provided monetary incentives to each of the participants at the end of the study.

After the rooms were chosen to set up the testbed, the third challenge was to interpret the readings from the sensors installed within the rooms. This is due to the fact that each user has different preferences and hence sets the sensors at different locations of the rooms. For instance, some may keep the sensors near the window whereby some prefer to keep them far from the windows. Similarly, some prefer to always keep their window curtains open whereas some prefer them to remain close most of the time. As a consequence, the readings on temperature, light, noise, and motion from the sensors have the possibility to be very different although they may represent the same environment. Hence, there is a need to consider such behavioral differences in our experiments, and we have tried to differentiate the sensors data based on the context. For example, we perform some sort of learning before interpreting the readings from sensors.

Finally, we find it difficult to match the theoretical load consumption model with the model that we derive from the experiments. For example, let us consider the case of air conditioning systems (ACSs). Although there are many studies on the energy consumptions of ACS in the literature, it is extremely difficult to find a study that has used a setup as same as the one we use, e.g., in terms of ACS type, room size and weather. Therefore, algorithms that are designed based on the theoretical model may not behave as expected in practical scenarios. As a result, additional steps are needed to build a model based on practical system setup, and building such model could be time consuming and very customized to a practical setting.

After discussing various challenges, now we will discuss the experiments that are currently (or will be) implemented at SUTD in the next section to give a glimpse of the competence of this extensive testbed setup.

\section{On-going Experiments}\label{sec:experiments} Most of the experiments using the testbed are deliberately related to electricity management either through direct control of electricity from the IES or by designing incentives that will encourage behavioral modification towards efficient energy use.  Our experiments at the SUTD testbed are conducted 1) within the student hostels, and 2) at the SUTD campus.  Based on their electricity usage pattern, the SUTD campus is considered as the commercial space in our experiments, whereby student hostels are thought of as residences.

\subsection{SUTD student hostels} The energy consumed by residences differ based on the set of home appliances, standard of living, climate, social awareness and residence type. We use our testbed to develop solutions for residential energy management to attain different objectives.
\subsubsection{Ancillary electricity markets}In the SUTD testbed, we are interested in determining the potential of residential smart grid participation in ancillary markets. Real-time ancillary markets allow participants to supply both regulation and reserve power.  Reserve power is generation capacity that is required during an unexpected outage of a scheduled generating plant and we are concerned with whether the residential users can provide the grid with either primary or secondary reserve power.
\begin{figure}[t!]
\centering
\includegraphics[width=\columnwidth]{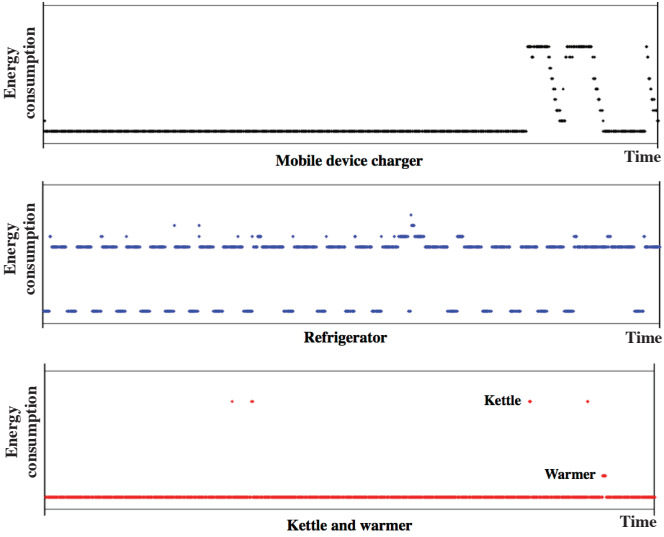}
\caption{Demonstration of the real-time energy consumption by different home appliances as obtained from the deployed testbed. The horizontal axis refer to different time of the day whereby the vertical axis represent the energy consumption by different appliances.} \label{fig:application}
\end{figure}
In most systems, discretionary loads can voluntarily participate in these markets as interruptible loads, willing to ``load-shed" in an exceptional event; however, current market rules only allow participation of big loads. In our system, the IES can aggregate smaller distributed loads and participate in these markets as a unified entity. Although theoretically possible, the practical implementation with consideration of practical limitations have not been fully examined in full scale testbeds, especially the communication constraints. With the SUTD testbed, we will verify its feasibility, and the requirements on the communication network in supporting such an application. 
\subsubsection{Dynamic pricing}Using the implemented testbed, we also focus on designing dynamic pricing schemes for residential users. Different design objectives such as penetration rates~\cite{Hassan-Energies:2013}, flexibility thresholds and flexibility constraints will be considered. As an initial step towards this development, a mobile application is being developed for the residential users at SUTD that can periodically notify them about their real-time energy consumption amount and the associated price per unit of energy. We will use such dynamic pricing information to achieve certain DRM goals and to promote energy saving awareness. The smart meter is installed in parallel with original meter, such that the smart meter can be used to simulate the electricity price based on dynamic pricing, while the original meter by the grid can continue provide meter readings for actual electricity billing. 
\subsubsection{DRM under communication impairment} Real-time DRM can be affected significantly by the quality of the communication network of the smart grid. In this context, we are conducting experiments to observe the effect of the status of communication network on DRM. For instance, if the communication network is congested, it would result in packet losses or delay. We are interested in investigating how this congestion of the network affects the DRM in terms of delay and consequently the cost, reduction of peak load and energy savings.
\subsubsection{DRM design based on preferences and device characteristics}The installed smart-plugs and MPNs in each HAN unit at the SUTD student hostel enables real-time monitoring of energy consumption of different appliances as shown in Fig~\ref{fig:application}. This provides insights on the students personal preferences and allows to feed the energy consumption data back to the consumers in real-time, which along with the incentive based DRM policies would enable the users to modify their usage pattern towards more efficient energy management.
\begin{remark}
It is important to note that our testbed is to monitor the energy usage. However, each appliance has a unique characteristic profile as shown in Fig.~\ref{fig:application}, which can be used to identify what appliance that is, and in return, speculate the activity and number of users.  This prompts the need for secure communication.
\end{remark}

\begin{figure}[t!]
\centering
\includegraphics[width=\columnwidth]{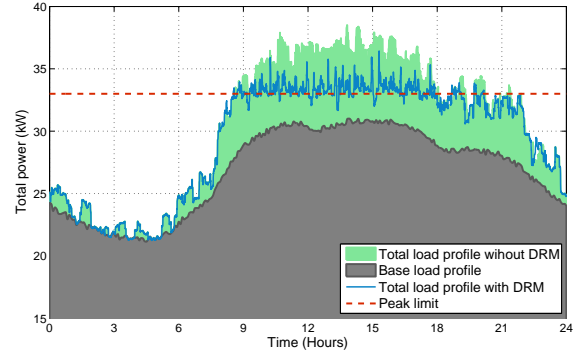}
\caption{Illustration of energy management scheme in the $10$ hostel units in order to reduce peak load.} \label{fig:peak_load_shave}
\end{figure}

Now to demonstrate how the proposed testbed can be used to perform energy management under practical communication impairment, we run experiment for peak load shaving in our testbed. To do so, we have considered $10$ home units where each unit is equipped with two flexible loads (we use lights as flexible loads that provide on/off control). We model the daily base load according to the reports of National Electricity Market of Singapore (NEMS) and scale according to the energy rating of light bulbs such that we can obtain a significant percentage of flexible load. We assume a threshold for total maximum allowable load consumption by all the units and design an algorithm to control the flexible load of each home units in order to keep the energy consumption always below the threshold. The algorithm is based on a centralized control scheme in which the energy management service provider conduct demand management by switching off several appliances of the users to maintain the total power demands below the given threshold. As designed, the energy management service provider considered not only the engagement of users in the demand response but also the inconvenience of users when the demand response protocol is conducted. We find that the demand response can be affected by communication delay. The detail of the algorithm and assumption of the experiment can be found in \cite{WenTai_Access:2015}.

The results from the experiment are demonstrated in Fig.~\ref{fig:peak_load_shave}. In Fig.~\ref{fig:peak_load_shave}, the green zone and blue zone of the figure denote the load profile with and without demand management respectively whereby grey zone indicate the base load and red line show the maximum allowable peak demand threshold (which is $33$ kW in the considered case). Now, as demonstrated in Fig.~\ref{fig:peak_load_shave}, when the total demand is under peak limit, there is no need to do any controlling and hence no difference is observed between green zone and the blue line. However, once the total demand exceed the peak limit, the algorithm is executed and thus controls the flexible loads in each home unit and turns some of them off to reduce demand load promptly as indicated by the blue line. It is important to note that the algorithm continuously run in the backend and monitors for situations when the total demand may exceed the threshold. Once such situation arises, the algorithm switches off some of the appliances to keep the demand within the threshold. However, the algorithm, as it is designed, considers not only the demand control but also the associated inconvenience that the users may experience for such control. Therefore, control of appliances is always kept at as low as possible in maintaining the demand. Furthermore, while controlling some appliances for handling excess demand, there could be new loads switched on by the users that can also contribute to the overall demand. As a consequence, there is no sudden decrease is observed in the blue line as can be seen from Fig. 4. Thus, the result in Fig.~\ref{fig:peak_load_shave} clearly shows that our developed testbed is effective to preform energy management applications. 

\begin{table}[t!]
\centering
\caption{Demonstration of how the monitoring of users occupancy and energy use of lights and ACS in eight office rooms at SUTD are done through the implemented testbed in order to reduce energy wastage. The total wastage is calculated in terms of both kWh and SGD.} 
\includegraphics[width=\columnwidth]{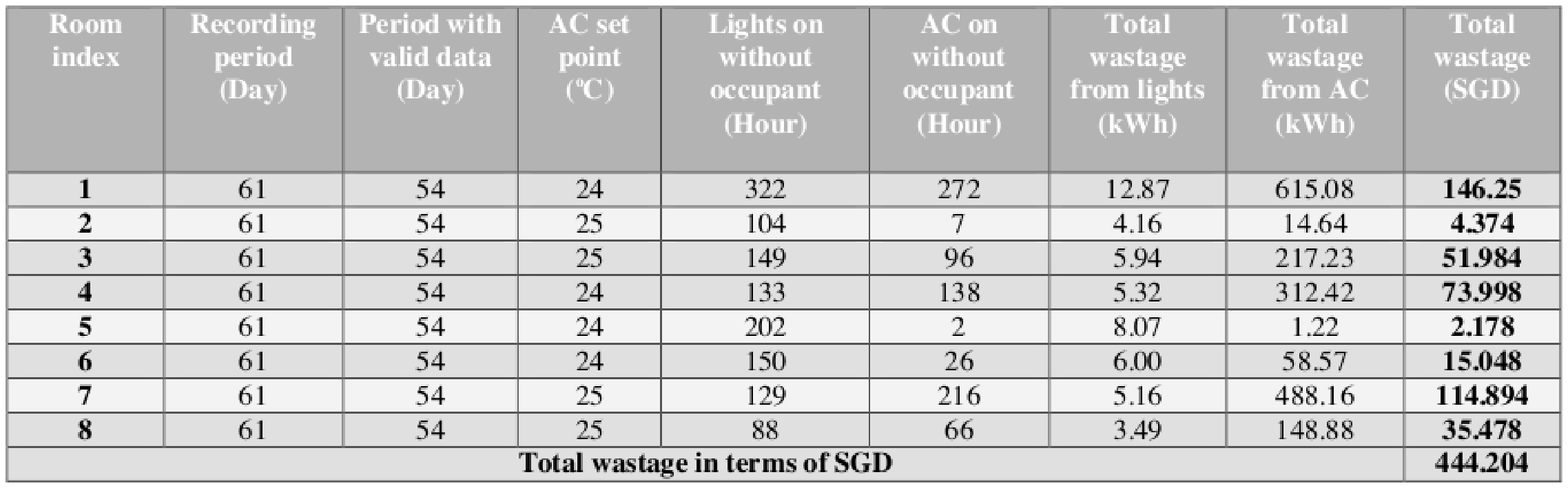}
\label{table:office}
\end{table}

\subsection{SUTD campus}
We are also interested in studying energy management in commercial buildings with a view to reduce energy costs for all stakeholders. To this end, we are currently conducting the following experiments at the SUTD campus.
\subsubsection{DRM in office environments}One of our ongoing experiments is to implement DRM schemes for offices that considers the real-time control of ACSs. The objectives are mainly two-fold: 1) real-time thermostat control to manage peak demand, and 2) optimal management of ACS demand under dynamic pricing for energy cost management. To support this, we are developing energy consumption models and real-time control algorithms for ACSs. The testbed allows us to monitor and collect  the data on consumed power by the ACSs under different permutations of indoor and outdoor conditions. These are pivotal for algorithm development and ACS energy consumption modeling. 
\subsubsection{Meeting scheduling} In offices, usually there exists a large number of meeting rooms that are used only occasionally. Hence, the scheduling of their usage is a possible way to shape the energy consumption of commercial buildings. We formulate a comprehensive study, where for a given set of meeting room requests over a fixed time period and a set of meeting rooms in a dynamic pricing market, the meeting scheduler finds a feasible routine that minimizes the total energy consumption (and/or cost). The SUTD testbed is used for obtaining the real-data, and to test the scheduler in order to verify the effectiveness of the optimal meeting scheduling protocol.

Now, in order to save energy and associated costs of individual office rooms, we experiment to identify the potential wastage in an office environment. To do so, we have deployed a number of sensors, i.e., the MPN of Fig.~\ref{fig:HAN} in each of the offices. For instance, if the sensors do not detect any motion or noise in a room for a predefined period of time, we can assume that there is no one inside the room. A record of such observation is shown in Table~\ref{table:office}, which shows the duration of turned-on lights and ACSs of eight different office rooms when there are no occupants. According to Table~\ref{table:office}, the energy wastage in most of the considered rooms is significant in absence of the occupants, which is on average $6.376$ kWh and $232.025$ kWh per room from lights and ACS respectively. Now, if we assume that the average wastage is the same for all 200 office rooms of a typical medium size university campus, the total energy that wasted during the considered time duration of 54 days can be estimated as $47,680.2$ kWh. Now considering the electricity rate in Singapore, which is calculated via Home Electricity Audit Form available online\footnote{The cost can be calculated by using the \emph{Home Electricity Audit Form} available online for public use in https://services. spservices.sg/elect\_audit\_frameset.asp.}, this wastage can be translated into a total of around $11,100$ SGD from the considered 200 rooms during a period of $54$ days. Therefore, if the proposed testbed can be used in offices for the considered setting, about $11,100$ SGD can be saved from wastage. The saving will be even more if the electricity price is dynamic, as the office hour is typically overlap with day time peak period. One can then perform on/off control to the ACSs (such as in Fig.~\ref{fig:peak_load_shave}) to achieve energy saving or peak load shaving.

\section{Conclusion}\label{sec:conclusion} In this paper, we have discussed some aspects on the development of the smart grid testbed at the Singapore University of Technology and Design (SUTD). These testbeds within the university campus have been setup to approximate both residential and commercial spaces. We have discussed the general components, features and related challenges of such a testbed implementation with emphasis on the creation of the  communication system architecture. Ongoing experiments at the testbed include the harnessing of detailed data streams from the testbed for different customized energy management programs such as demand response. The flexibility and extensivity of the deployed testbed allows for the research team to explore and implement effective and practical communications technologies and smart grid applications, which could ultimately increase the acceptance and adoption rate of these systems.
\section*{Acknowledgements}
This work is supported in part by the Singapore University of Technology and Design through the Energy Innovation Research Program Singapore under Grant NRF2012EWT-EIRP002-045, in part by the SUTD-MIT International Design Center, Singapore under Grant IDG31500106 and in part by NSFC-61550110244.


\begin{thebibliography}{15}
\providecommand{\url}[1]{#1}
\csname url@samestyle\endcsname
\providecommand{\newblock}{\relax}
\providecommand{\bibinfo}[2]{#2}
\providecommand{\BIBentrySTDinterwordspacing}{\spaceskip=0pt\relax}
\providecommand{\BIBentryALTinterwordstretchfactor}{4}
\providecommand{\BIBentryALTinterwordspacing}{\spaceskip=\fontdimen2\font plus
\BIBentryALTinterwordstretchfactor\fontdimen3\font minus
  \fontdimen4\font\relax}
\providecommand{\BIBforeignlanguage}[2]{{%
\expandafter\ifx\csname l@#1\endcsname\relax
\typeout{** WARNING: IEEEtran.bst: No hyphenation pattern has been}%
\typeout{** loaded for the language `#1'. Using the pattern for}%
\typeout{** the default language instead.}%
\else
\language=\csname l@#1\endcsname
\fi
#2}}
\providecommand{\BIBdecl}{\relax}
\BIBdecl

\bibitem{Fang-J-CST:2012}
X.~Fang, S.~Misra, G.~Xue, and D.~Yang, ``Smart grid - {T}he new and improved
  power grid: {A} survey,'' \emph{IEEE Commun. Surveys Tuts.}, vol.~14, no.~4,
  pp. 944--980, Oct 2012.

\bibitem{Maharjan-TSG:2013}
S.~Maharjan, Q.~Zhu, Y.~Zhang, S.~Gjessing, and T.~Ba{\c{s}}ar, ``Dependable
  demand response management in the smart grid: {A} {S}tackelberg game
  approach,'' \emph{IEEE Trans. Smart Grid}, vol.~4, no.~1, pp. 120--132, Mar
  2013.

\bibitem{Rong-Networks:2011}
R.~Yu, Y.~Zhang, S.~Gjessing, C.~Yuen, S.~Xie, and M.~Guizani, ``Cognitive
  radio based hierarchical communications infrastructure for smart grid,''
  \emph{IEEE Netw.}, vol.~25, no.~5, pp. 6--14, Oct 2011.

\bibitem{Wayes_ITS:2016}
W.~Tushar, C.~Yuen, S.~Huang, D.~Smith, and H.~V. Poor, ``{Cost minimization of
  charging stations with photovoltaic: An approach with EV classification},''
  \emph{IEEE Trans. Intell. Transp. Syst.}, vol.~17, no.~1, pp. 156--169, Jan.
  2016.

\bibitem{Naveed_Access:2015}
A.~Naeem, A.~Shabbir, N.~U. Hassan, C.~Yuen, A.~Ahmed, and W.~Tushar,
  ``{Understanding customer behavior in multi-tier demand response management
  program},'' \emph{IEEE Access (Special issue on Smart Grids: A Hub of
  Interdisciplinary Research)}, vol.~3, pp. 2613--2625, Nov. 2015.

\bibitem{Atzeni-TSG:2013}
I.~Atzeni, L.~G. Ord{\'{o}}{\~{n}}ez, G.~Scutari, D.~P. Palomar, and J.~R.
  Fonollosa, ``Demand-side management via distributed energy generation and
  storage optimization,'' \emph{IEEE Trans. Smart Grid}, vol.~4, no.~2, pp.
  866--876, June 2013.

\bibitem{huang2012effects}
Y.~Liu, C.~Yuen, S.~Huang, N.~Ul~Hassan, X.~Wang, and S.~Xie, ``Peak-to-average
  ratio constrained demand-side management with consumer's preference in
  residential smart grid,'' \emph{IEEE J. Sel. Topics Signal Process.}, vol.~8,
  no.~6, pp. 1084--1097, Dec 2014.

\bibitem{Hassan-Energies:2013}
N.~U. Hassan, M.~A. Pasha, C.~Yuen, S.~Huang, and X.~Wang, ``Impact of
  scheduling flexibility on demand profile flatness and user inconvenience in
  residential smart grid system,'' \emph{Energies}, vol.~6, no.~12, pp.
  6608--6635, Dec 2012.

\bibitem{TaoJiang_TSG:2014}
T.~Jiang, Y.~Cao, L.~Yu, and Z.~Wang, ``Load shaping strategy based on energy
  storage and dynamic pricing in smart grid,'' \emph{IEEE Trans. Smart Grid},
  vol.~5, no.~6, pp. 2868--2876, Nov 2014.

\bibitem{LiangYu_TPDS:2015}
L.~Yu, T.~Jiang, and Y.~Cao, ``Energy cost minimization for distributed
  internet data centers in smart microgrids considering power outages,''
  \emph{IEEE Trans. Parallel Distrib. Syst.}, vol.~26, no.~1, pp. 120--130, Jan
  2015.

\bibitem{Tushar-TIE:2015}
W.~Tushar, B.~Chai, C.~Yuen, D.~B. Smith, K.~L. Wood, Z.~Yang, and H.~V. Poor,
  ``Three-party energy management with distributed energy resources in smart
  grid,'' \emph{IEEE Trans. Ind. Electron.}, vol.~62, no.~4, pp. 2487--2498,
  Apr. 2015.

\bibitem{Liu-JIEEEIT:2014}
Y.~Liu, C.~Yuen, X.~Cao, N.~Hassan, and J.~Chen, ``Design of a scalable hybrid
  mac protocol for heterogeneous {M2M} networks,'' \emph{IEEE Internet Things
  J.}, vol.~1, no.~1, pp. 99--111, Feb 2014.

\bibitem{Tushar-TSG:2014}
W.~Tushar, J.~A. Zhang, D.~B. Smith, H.~V. Poor, and S.~Thi{\'{e}}baux,
  ``Prioritizing consumers in smart grid: {A} game theoretic approach,''
  \emph{IEEE Trans. Smart Grid}, vol.~5, no.~3, pp. 1429--1438, May 2014.

\bibitem{Cao:2016}
Y.~Cao, T.~Jiang, M.~He, and J.~Zhang, ``Device-to-device communications for
  energy management: {A} smart grid case,'' \emph{IEEE J. Sel. Areas Commun.},
  vol.~34, no.~1, pp. 190--201, 2016.

\bibitem{WenTai_Access:2015}
W.-T. Li, C.~Yuen, N.~U. Hassan, W.~Tushar, and C.-K. Wen, ``{Demand response
  management for residential smart grid: From theory to practice},'' \emph{IEEE
  Access (Special issue on Smart Grids: A Hub of Interdisciplinary Research)},
  vol.~3, pp. 2431--2440, Nov. 2015.

\end{thebibliography}
\end{document}